# Population Fitness and Genetic Load of Single Nucleotide Polymorphisms Affecting mRNA splicing

Peter K Rogan[1,2*] and Eliseos J Mucaki[2]

Departments of [1]Biochemistry and [2]Computer Science, University of Western Ontario, London, Canada N6A 5C1.

**ABSTRACT**

Deleterious genetic variants can be evaluated as quantitative traits using information theory-based sequence analysis of recognition sites. To assess the effect of such variants, fitness and genetic load of SNPs which alter binding site affinity are derived from changes in individual information and allele frequencies. Human SNPs that alter mRNA splicing are partitioned according to their genetic load. SNPs with high genetic loads (>0.5) are common in the genome and, in many instances, predicted effects are supported by gene expression studies.

## 1   INTRODUCTION

The identification of true risk alleles in genome wide association studies has often been confounded by population structure. Linkage disequilibrium can extend over large genomic distances and depends on chromosome geneology, making it challenging to pinpoint causative alleles (Kimmel et al. 2008). High density genomic sequencing accompanied by computational approaches to predict deleterious variants may potentially overcome the challenges posed in some GWAS analyses. These methods aim to prioritize analysis of SNPs to those with functional effects on gene products, gene structure and/or expression.

The effects of natural sequence variation at nucleic acid binding sites can be assessed with models based on Shannon information, since individual information content is related to the affinity of the binding event (Schneider 1997). Phenotypes can be predicted from corresponding changes in the individual information contents ($R_i$, in bits) of the natural and variant DNA binding sites (Rogan *et al.*, 1998). In mRNA splicing, mutant sites have lower $R_i$ values than the corresponding natural sites, with null alleles having values at or below 1.6 bits (Rogan et al. 2003; Kodolitsch *et al.,* 1999). The decreased $R_i$ values of mutated splice sites indicate that such sites are either not recognized or are bound with lower affinity, and may result in exon skipping or cryptic site activation. Decreases in $R_i$ are more moderate for leaky splicing mutations and are associated with milder phenotypes. The minimum change in binding affinity for leaky mutations is $\geq 2^{\Delta Ri}$ lower fold than cognate wild type sites. Mutations that activate cryptic splicing may decrease the $R_i$ value of the natural site, increase the strength of the cryptic site, or concomitantly affect the strengths of both types of sites (see Figure 2). Non-deleterious changes result in negligible changes in $R_i$ (Rogan and Schneider, 1995). Increases in $R_i$ indicate stronger interactions between protein and cognate binding sites.

The information content of a sequence measures the success (or fitness) of the organism in its environment (Adami et al 2000). Neutral mutations which change Shannon entropy are critical for evolution to proceed (Maynard-Smith 1970). Adaptive evolution can be viewed (in the absence of frequency-based selection) as the maximization of fitness in a population. The fittest genotype is the one with the highest growth rate. Expressed in an information theory framework, growth rate (or channel capacity) is limited by the length of the message encoding the genotype, the bit rate of transmission from parent to offspring and the power to noise ratio, which is equivalent to robustness to effects of mutation.

In this study, we develop a fitness metric that detects deleterious alleles based on changes in information content at functional binding sites. By determining their mutational load, we apply this measure to detect splicing mutations in the population resulting from validated single nucleotide polymorphisms (SNPs). Many of these predictions can be confirmed with gene expression data.

## 2   METHODS

### 2.1   Objective function

The objective function relates differences in individual Shannon information content of a pair of functional alleles ($R_{i,major}$, $R_{i,minor}$) to the fitness of these genotypes in a population of individuals. Fitness is a usually unitless measurement of replication ability of a particular individual in a specified environment. To apply fitness (W) in the context of the information content of a pair alleles, we assume a homogeneous genetic landscape in the absence of any other data regarding their origin. In population genetics, comparison of fitnesses of different genotypes gives a ratio of their respective replication rates. Fitness is defined as the log of the actual replication rate, because merits of one genotype relative to another are doled out using an exponential scheme. Replication rate is the merit divided by the mean gestation time (assumed to be constant for genotypes exposed to the same environment). In artificial organisms merit increases exponentially with reproductive success, and the information content corresponds to the log of its fitness (Adami 1998).

$$R_i \cong \log_2 W \qquad (1)$$

*To whom correspondence should be addressed at progan@uwo.ca

Rearranging eq (1),

$$W \cong 2^{R_i} \quad (2)$$

The difference in the individual information contents of the minor and dominant alleles ($R_{i,major} - R_{i,minor} = \Delta R_i$) can be restated in terms of the fitness of these alleles.

For a single gene locus with the alleles $A_1..A_n$, which have the fitnesses, $w_1..w_n$ and the allele frequencies, $p_1..p_n$, respectively, $\bar{w}$ is the mean of all the fitnesses weighted by their corresponding frequency of allele $A_i$:

$$\bar{w} = \sum_{i=1}^{n} p_i w_i \quad (3)$$

Expressing mean fitness in terms of the individual information of a SNP with two alleles:

$$\bar{w} = p_1 \, 2^{R_{i,major}} + p_2 \, 2^{R_{i,minor}} \quad (4)$$

Ignoring frequency-dependent selection, then locus-specific genetic load ($L$) is:

$$L = \frac{w_{max} - \bar{w}}{w_{max}} \quad (5)$$

where $w_{max}$ is the maximum possible fitness, $2^{R_{i,major}}$. The multi-locus mean fitness (in the absence of linkage between loci) is the product of the genetic loads for individual loci (Crow and Kimura 1970):

$$\bar{w}_{POPULATION} = \prod_{j=1,n} (1 - L_j) \quad (6)$$

### 2.2 Information and expression analysis of HapMap SNPs

SNP-induced changes in natural splice site information were detected by scanning all human chromosome sequences with information weight matrices based on sequences of known donor and acceptor splice regions from both strands of the human genome reference sequence (Rogan *et al.*, 2003). The average information contents of the acceptor and donor sites in these models are respectively, $9.8118 \pm 0.0001$ bits/site and $8.12140 \pm 0.00009$ bits/site. Splicing related changes in the information contents of splice donor and acceptor sites were computed for all SNPs in the reference human genome (hg18) that have been validated in the Phase II human haplotype map (www.hapmap.org). The results were filtered to select changes in Ri values of all annotated donor and acceptor sites. These changes have been organized and related to known gene structures, population genotypes and allele frequencies and gene expression results.

We created a MySql database to integrate microarray and EST expression data with gene structure, SNP genotype, allele frequency and changes in splice site information. The gene expression data are derived from Genbank (mRNAs and ESTs) and from GEO dataset (GSE7792; Huang *et al*, 2007), which covers all exons on the Affymetrix exon ST array hybridized with cDNAs of genotyped HapMap samples. Exon expression data was derived from Affymetrix Human Exon 1.0 ST microarray data of 176 genotyped HapMap cell lines. Probeset intensities were re-computed by eliminating probes which overlapped any known SNP (dbSNP129) prior to normalization. Microarray data was normalized using the PLIER algorithm, then incorporated into a MySQL database with tables containing HapMap genotypes and Ensembl gene and exon boundaries. The database was used to relate SNPs to their nearest probesets to find genotype-related changes in the splicing index (defined as probeset intensity/overall gene intensity) of an exon. SNPs found within donor and acceptor splice sites where genotype-specific splicing index changes (major homozygote, heterozygote, minor homozygote genotypes where possible) are present in a related probeset.

SQL queries were used to identify changes in gene expression concordant with predictions of information theory based changes due to SNPs within naturally-occuring donor and acceptor splice sites. We have previously shown that changes in information content can be due to SNP-related splicing mutations and that these changes affect gene expression (Nalla *et al.*, 2005).

Perl scripts were developed to compute genetic fitness and load for all SNPs based on information-based changes in mRNA splice sites and population allele frequencies. Analyses were performed both on a composite set of CEU (Caucasian; n=90) and YRI (Yoruba/Nigerian; n=90) individuals. Offspring of CEU trios were removed from the allele frequency calculations. Each population was also analyzed separately for intergroup comparisons of the effects of differences in allele frequencies.

## 3 RESULTS

### 3.1 Fitness and genetic load estimation

Both $R_i$ values and allele frequencies (Table 1) affect genetic load and fitness as predicted. Rare SNPs ($p_2 = 0.01$) with small changes in Ri (<0.5 bit) marginally reduce genetic load and population fitness. Since this corresponds to ~1.4 fold lower binding site affinity, there is only a modest change in expression due to such SNPs. The multilocus mean fitness is reduced for such mutations, but these variants can still be tolerated at numerous genomic loci. By contrast, low frequency alleles ($0.05<p_2<0.1$) with larger $\Delta R_i$ values (1 to 5 bits) depress the load and population mean fitness to unsustainable levels when present at many loci (n=100).

**Table 1**. Fitness and load computation using simulated parameters

| $R_{i,\,major}$ | $R_{i,minor}$ | $p_2$ | $\bar{w}$ | $w_{max}$ | $L$ | # Loci | $\bar{w}_{POP}$ |
|---|---|---|---|---|---|---|---|
| 10 | 5 | 0.1 | 924 | 1024 | 0.09 | 1 | 0.9 |
| 10 | 5 | 1.0 | 32 | 1024 | 0.97 | 1 | 0.03 |
| 8 | 7 | 0.1 | 243 | 256 | 0.05 | 1 | 0.95 |
| 10 | 5 | 0.1 | 924 | 1024 | 0.09 | 10 | 0.36 |
| 10 | 5 | 0.1 | 924 | 1024 | 0.09 | 100 | 0.00004 |
| 8 | 7 | 0.1 | 243 | 256 | 0.05 | 10 | 0.6 |
| 8 | 7 | 0.1 | 243 | 256 | 0.05 | 100 | 0.01 |
| 8 | 7.5 | 0.01 | 252 | 256 | 0.01 | 100 | 0.18 |

Interestingly, an increase in Ri of a rare minor allele can significantly increase the genetic load. In these instances, selection is operating to reduce exon recognition, and increase exon skipping.

### 3.2 Distribution of SNP-related changes in individual information at natural splice sites

*3.2.1 ΔR$_i$ values at natural splice sites.* The human genome reference sequence was analyzed for all information changes due to Phase II HapMap SNPs (n=4,071,589). The information contents of cryptic and natural splice sites within known genes were altered for 1,093,474 of these variants (361,003 donor sites, 732,471 acceptor sites, which includes cryptic splice sites). Of these, 9051 SNPs altered the $R_i$ values of natural donor and acceptor sites in 5970 genes, with 211 having more than one SNP mapped to the same site. There were 8 instances where two different natural splice sites were affected by the same SNP. The information changes affected similar numbers of SNPs for both minor and major alles ($R_{i,major}$ > $R_{i,minor}$: 4311; $R_{i,minor}$ > $R_{i,major}$: 4740). As expected from selection, the quantity natural sites altered by SNPs is related to the severity of the mutation (Table 2).

**Table 2.** Splice sites affected by SNPs for different ΔR$_i$ thresholds

| ΔR$_i$ (bits) | Total # sites | # Donor | # Acceptor |
|---|---|---|---|
| ≥ 0.1 | 8873 | 1767 | 7106 |
| ≥0.5 | 7064 | 1347 | 5717 |
| ≥1.0 | 3097 | 963 | 2134 |
| ≥5.0 | 462 | 145 | 317 |

### 3.3 Fitness and genetic loads from combined HapMap populations

*3.3.1 Intrasite distribution of sequence changes.* Position-specific substitutions within natural splice donor and acceptor sites were analyzed for rare and common SNPs with predicted effects on splicing. SNPs with small load values (<0.12) were filtered in order to minimize variants with limited effects on splicing (likely the result of genetic drift). Substitutions in the conserved dinucleotide (0, +1) adjacent to all splice junctions are significantly less common than changes elsewhere (some common SNPs do occur at these positions). In donor sites, the patterns of substitution are consistent at low and high allele frequencies. Overall, substitutions due to rare SNP alleles are ~5 fold less abundant than for common alleles. Substitutions in protein coding sequences frequently alter strengths of both donor (-1 to -3) and acceptor (+1, +2) splice sites.

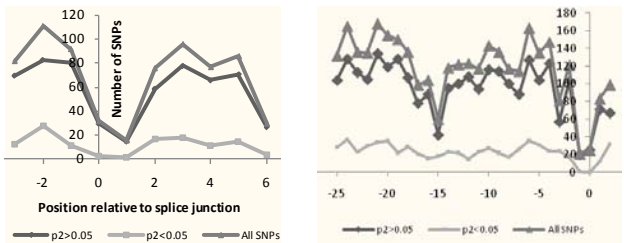

**Figure 1.** Position-specific substitutions due to SNPs in donor (left) and acceptor (right) splice sites with genetic loads ≥ 0.12.

.
*3.3.2 Genetic load analysis.* SNPs with L> 0 occuring at natural splice sites in 5374 different genes (Table 3). The

**Table 3.** Relationship between genetic load and average ΔR$_i$

| L > | No. Sites | $\overline{\Delta R_i}$ (bits) | ±Std.Dev. |
|---|---|---|---|
| 0.00 | 7476 | | |
| 0.10 | 5375 | 0.36 | 2.4 |
| 0.12 | 5046 | 0.44 | 2.4 |
| 0.20 | 3773 | 0.82 | 2.34 |
| 0.30 | 2634 | 1.17 | 2.44 |
| 0.40 | 1426 | 1.84 | 2.73 |
| 0.50 | 779 | 2.69 | 3.07 |
| 0.60 | 484 | 3.19 | 3.43 |
| 0.70 | 293 | 3.72 | 3.78 |
| 0.80 | 146 | 4.94 | 4.75 |

majority of these variants with low loads (L<0.12) exhibit small changes in information (ΔR$_i$ <1 bit), limiting their impact on splicesomal binding affinity and exon recognition. The contribution of the ΔR$_i$ term predominates over allele frequency at larger loads (L>0.5), because of larger difference between the mean and maximum fitness. Sensitive quantitative RT-PCR assays do not consistently detect changes in transcript levels below ΔR$_i$<0.5 bits (Rogan and Mucaki, in preparation). At low load values, small effects due to segregating SNPs that result from genetic drift are challenging to sort out from phenotypically significant modest changes in mRNA splicing. The relationship between L and $\overline{\Delta R_i}$ is not a simple one, which is evident from the large standard deviations on

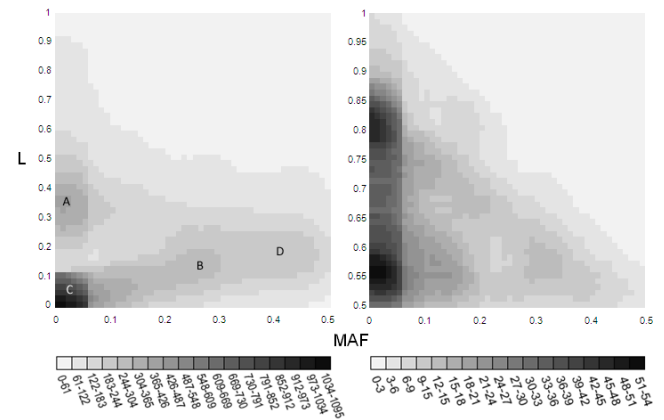

**Figure 2.** Genetic loads (L) of SNPs in CEU + YRI populations sorted by minor allele frequency (MAF). Panels depict all SNPs across full range of loads (0 – 1; left), and exclusively those with high loads (0.5 – 1; right). Legends below the corresponding panel indicate graduated grayscale distributions of SNP densities, each thresholded by a range of counts. The plot resolution of each window is 0.01 L x 0.01 MAF units. Labels A-D depict locally dense concentration of SNPs.

$\overline{\Delta R_i}$.

The genome-wide distribution of high load SNPs affecting mRNA splicing was partitioned according to allele frequency (Figure 2). Rare MAF alleles span the full range of genetic loads (0 – 0.98), whereas L for common alleles is highly constrained (≤0.5), probably because of selection. Based on results in Tables 2 and 3, we expected to find fewer SNPs with higher loads with increasing ΔR$_i$ values due to selection. However, the SNP density plots (Fig. 2) of L over the full range of minor allele frequencies reveals a more complex distribution.

The distribution of SNPs with rare alleles ($p_2 \leq 0.05$) is multimodal, with major peak concentrations from $0 < L \leq 0.11$ (region A) and $0.2 \leq L \leq 0.5$ (region C). This partitioning may prove to have functional significance. Region A consists of 267 SNPs, with $0.51 \leq \Delta R_i \leq 0.78$ bits. Interestingly, the $R_i$ values of the minor alleles of *all* splice sites altered by these SNPs are stronger than the corresponding major alleles. Region C contains 930 SNPs, of which 696 have minor alleles with lower $R_i$ values (-18.63 $\leq \Delta R_i \leq$ -0.5; $\overline{\Delta R_i}$ = -1.55 ± 2.81 bits). Few these SNPs have increased $R_i$ values, with a maximum of 0.2 bits.

Locally dense pockets of common SNPs are clustered within a narrow range of low load values (regions B [$0.1 \leq L \leq 0.15$] and D [$0.1 \leq L \leq 0.2$]) and a broad range of allele frequencies. Region B consists of 239 SNPs with minor alleles with predominantly lower $R_i$ values (-1.7 $\leq \Delta R_i \leq$ 0.35; $\overline{\Delta R_i}$ = -0.70 ± 0.42 bits). Stratifying this region, SNPs with higher loads (0.13-0.15) exhibit larger reductions in $\Delta R_i$ (> -1 bits). Region D consists of a block of 433 SNPs, most of which show limited changes in $R_i$ values (-1.3 $\leq \Delta R_i \leq$ 0.66; $\overline{\Delta R_i}$ = 0.22 ± 0.59 bits). Only 18 splice sites have SNPs with $\Delta R_i \geq$ -0.9 bits.

*3.3.3 Gene expression validation of predicted SNP effects.*

Predictions of information analysis were compared with microarray expression data for exons associated with these SNPs. Initially, splicing indices (SI) of relevant probesets were simply compared with $\Delta R_i$ values of splice sites associated with SNPs predicted to reduce splice site use. There are limitations with the accuracy of this approach, as probesets often can show considerable variation in SI values in individuals with the same genotypes. SNPs (n=355) that concomitantly activate one or more adjacent cryptic sites within the same probeset were also eliminated from the analysis, since these would not be expected to alter SI.

**Table 4**. Relationships between SI and $\Delta R_i$

| | Range of SI ratios between: | | | |
|---|---|---|---|---|
| $\Delta R_i$(maj-min) | 0.7 and 0.9 | 0.5 and 0.7 | >0 and 0.5 | Total # SNPs |
| > 2 | 47 | 8 | 11 | 66 |
| 1 to 2 | 70 | 26 | 13 | 109 |
| 0.5 to 1 | 251 | 84 | 75 | 410 |
| 0 to 0.5 | 102 | 28 | 21 | 151 |
| -0.5 to 0 | 94 | 20 | 12 | 126 |
| -0.5 to -1 | 191 | 54 | 61 | 306 |
| -1 to -2 | 83 | 32 | 38 | 153 |
| < -2 | 83 | 39 | 43 | 165 |

We anticipated that categories of SNPs with higher $\Delta R_i$ values would be associated with lower SI expression changes. While there is some suggestion that this may be correct (SI: compare >0 - 0.5 vs 0.7 – 0.9), the results are not compelling (Table 4). Statistical tests developed to predict alternative splicing events from microarry expression data have also had mixed results (Hu et al. 2001; Wang et al. 2003).

The use of high load SNPs with allele-specific, dose-dependent expression of the weaker splice site effectively eliminates SNPs with high coefficients of variation of interindividual expression and permits direct analysis of allele specific effects. We analyzed 1069 SNPs in which hetero- and homozygous individuals carrying the lower Ri value allele were concordant for SI (either both >1 or both <1). For example, rs3764897 is a common SNP (MAF= 0.23, L=0.634) in exon 23 of the PLD2 gene, which strengthens an alternative donor site downstream (from G: 1.5 → A: 4.1 bits) leading to a 33 nt extension of this exon (Figure 3). The median SI of probeset 3707250 which covers the extended exon is consistent with increasing dosage of the A allele.

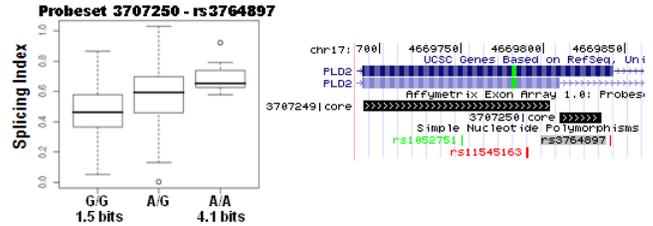

**Figure 3.** Expression changes at a common, high-load SNP. Whisker plot indicates dose-dependent expression in HapMap individuals of probeset 3707250 for all genotypes. Browser track shows the locations of rs3764897 in *PLD2*, the alternative splice site affected and probeset 3707250.

rs924900 is an example of a rare, high-load SNP (MAF=0.02; L=0.74) that affects inclusion of the adjacent exon in *ZNF713* (Figure 4). The G allele weakens the acceptor site ($\Delta R_i$=2 bits), reducing its expression and increasing exon skipping of the adjacent exon.

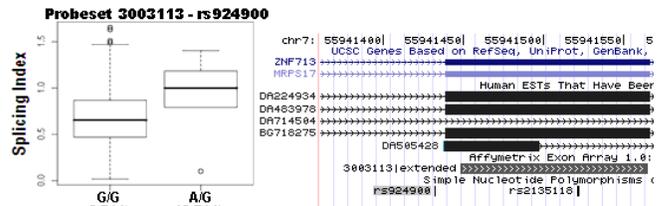

**Figure 4.** Expression changes at a rare, high-load SNP. Whisker plot indicates expression in HapMap individuals of probeset 3003113, which detects an increase in expression in AG heterozygotes. Browser track shows locations for rs924900 in ZNF713, probeset 3003113, and evidence of exon skipping in GenBank Accession DA714504.

*3.3.4 Mean fitness for HapMap individuals*

The mean fitness of each of the mRNA splicing mutations for all high load SNPs was determined explicity for each of the HapMap individuals. Higher fitness SNP homozygous and heterozygous genotypes are assumed to be equivalent. Of the 9051 SNPs (L>0) which affect splice site strength, each of these individual are homozygous for, on average, 471±23 SNPs with the lower fitness alleles. Figure 5 shows representative distributions of $\overline{w}$ for samples of 30 individuals from each of these HapMap populations. In each individual, the majority of loci consist of slightly deleterious mutations with average $\overline{w} \sim$ 0.8. It is also evident, however, that every individual harbors between 5 and 15 deleterious splicing variants ($0.02 < \overline{w} \leq 0.56$). Regardless of which individual is analyzed, $\overline{w}_{POPULATION} \approx 0$, and it is therefore underestimated by Eqn (6). When mild mutations are eliminated from the analysis, and $\overline{w}_{POPULATION}$ is computed from the SNPs with low $\overline{w}$ values (in Figure 3), similar results are obtained.

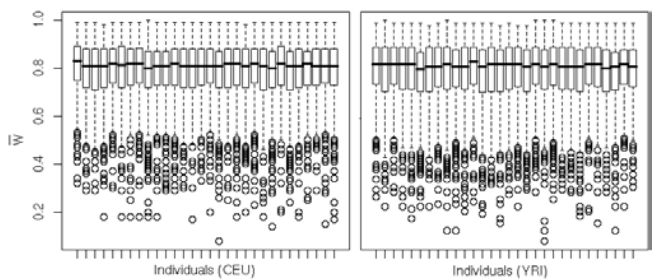

**Figure 5.** Distribution of mean fitness values of all SNPs affecting mRNA splicing in 30 CEU (left) and 30 YRI (right) individuals. Whisker plots indicate $\bar{w}$ for only homozygous genotypes.

**Fitness and genetic load of HapMap subpopulations**

*3.4.1 Overall comparison of SNP loads in CEU and YRI populations.* The effects of population specific selection were investigated by analyzing differences between genetic loads for the same SNP in different populations. The population allele frequencies of the CEU and YRI subgroups in Phase II HapMap data were used to compute population-specific genetic fitness and load values for each SNP. The analysis was limited to these groups because the corresponding exon microarray expression data were available for these groups only. Genetic drift contributions should be negligible because individual genotypes are sampled from large breeding populations (Li 1955).

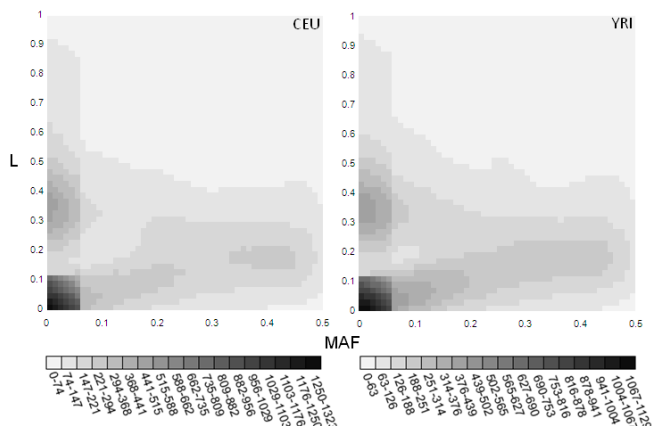

**Figure 6.** Genetic loads (L) of SNPs in natural splice sites for different populations. Panels show densities of SNPs in CEU (left) vs. YRI (right) groups sorted by minor allele frequency (MAF). Legends below each corresponding panel indicate graduated grayscale distributions of SNP densities, each thresholded by a range of counts

The overall patterns of SNP density within 9051 splice sites across a range of genetic loads are similar between the groups, especially for SNPs with MAF < 0.05. Nevertheless, 269 of these SNPs are polymorphic only in CEU and 152 are polymorphic only in the YRI group. Higher densities of common SNPs of moderate load are apparent in the YRI population; for example, there are increased number of SNPs with 0.08 <MAF< 0.16 at 0.22 <L< 0.35 and 0.25 < MAF <0.34 at 0.2 <L< 0.05.

Population-specific differences in frequencies of alleles can be discerned by comparing the loads for the same SNP in each subgroup, ΔL. Many SNPs have identical loads in both populations (ΔL=0; n=2715). Significant number of HapMap SNPs are either absent in CEU and YRI (n=2076) or are have very low MAFs in both of these populations (n=639).

**Table 5.** Population-specific differences in genetic load

| | Number of SNPs: | |
|---|---|---|
| ΔL≥ | YRI > CEU | CEU>YRI |
| 0.4 | 1 | 4 |
| 0.3 | 23 | 32 |
| 0.25 | 57 | 60 |
| 0.2 | 127 | 124 |
| 0.15 | 274 | 262 |
| 0.1 | 604 | 613 |
| 0.01 | 2557 | 2756 |

As expected, the quantity of SNPs decreases with increasing ΔL (Table 5). Larger values of ΔL presumably reflect bigger differences in the fitness landscapes between the populations. This metric is constrained by species specific selection and the divergence time since both groups shared a set of ancestral genotypes. Interestingly, the numbers of SNPs which show load differences are nearly balanced in the different populations, which is consistent with a fixed rate of frequency dependent selection of these SNPs in these respective populations since the migration of the CEU population from the African continent.

*3.4.2 Population specific differences in gene expression.*

Patterns of gene expression were also surveyed in SNPs exhibiting differences in genetic load due to frequency dependent selection. Dose-dependent, allele-specific changes in mRNA splicing were found for minor alleles of 637 SNPs present at different frequencies in CEU and in the YRI populations. Figures 7 and 8 show results for two such SNPs: rs10190751 which is more prevalent in YRI (ΔL =Δ$p_2$= 0.25) and rs1018448, which is more common in CEU (ΔL =Δ$p_2$= 0.36).

Information analysis of rs10190751 predicts ≥181 fold reduced affinity for the A allele in the acceptor site of exon 8 of CFLAR. Expression differences are significant for all genotypes (K-W $\chi^2$ =61, p<<0.001). Exon skipping due to decreased recognition of this acceptor is common in both mRNA and EST sequences.

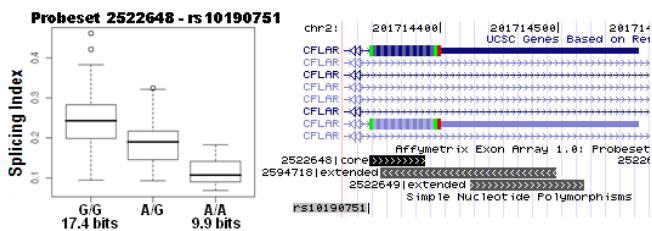

**Figure 7.** Expression changes at a high-load SNP present at high frequency in Yorubans and lower frequency in Caucasians. Whisker plot indicates expression in HapMap individuals of probeset 3707250 for all genotypes. Browser track shows the locations of rs10190751 in *CFLAR* and probeset 2522648, which activates an alternative terminal exon in individuals containing the G allele.

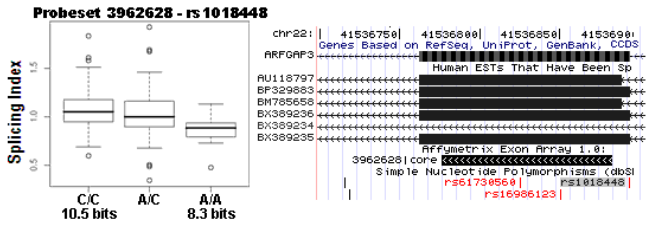

**Figure 8.** Expression changes at a high-load SNP present at high frequency in Caucasions and lower frequency in Yorubans. Whisker plot indicates expression in HapMap individuals of probeset 3292628, which shows graduated reduction in expression of this sequence with increasing copy number of the A allele. Browser track shows the locations of rs1018448 in exon 12 of *ARFGAP3*. GenBank Accession BX389234 lacks this exon.

A C-A substitution due to rs1018448A at a acceptor site of exon 12 in the *ARFGAP3* gene results in a 2.2 bit weaker binding site ($\geq 4.6$ fold, reducing inclusion of this exon in the mature mRNA, and may lead to increased skipping of this exon.

## 4  DISCUSSION

To date, the common variants described by genomewide association studies have had modest predictive power (generally, <10%). For most diseases and traits, these variants explain only a small fraction of heritability (Hirschorn 2009). Hundreds of risk alleles, each with distinct locus associations, are needed to explain the relative risks for complex genetic diseases (Kraft and Hunter 2009). Many rare alleles which contribute to these risks remain to be discovered. The present study suggests that the human genome contains *numerous* common and rare alleles with significant genetic loads. Many of these predicted mutations demonstrate dose-dependent, allele-specific effects on the exons adjacent to the splice sites in which they are found. A significant number of SNPs that alter mRNA splicing are under frequency-dependent selection in different human populations. We also validate these findings with corresponding gene expression data.

The proportion of mutations that are due to adaptive evolution has been estimated for protein coding changes, however these estimates is generally biased downward by segregation of slightly deleterious variants. To reduce this bias, low frequency polymorphisms have been removed (Charlesworth and Eyre-Walker 2008). Caution must be exercised when eliminating rare splicing mutations, where rare alleles are may not be slightly deleterious and can significantly impact mRNA structure and protein coding.

Each HapMap individual carries higher load genotypes, most of which are mild mutations with high $\overline{W}$ values. The multilocus mean fitnesses based on all of these variants are underestimated by equation (6). All individuals possess generally 5-15 low fitness outlier genotypes with low values of $\overline{W}$. Even if the SNPs with high $\overline{W}$ values are eliminated, these outlier loci dramatically depress $\overline{w}_{POPULATION}$ to unrealistically low levels. Accurate determination of multilocus mean fitness is challenging for a variety of reasons. The combined deleterious effects of mutations in proteins are subject to negative epistasis which are larger than expected from the multiplication of their individual effects (Bershtein et al., 2006). The unobserved geneology of ancestral recombination patterns invalidate assumptions regarding the independence of individual fitness scores (Griffiths and Marjoram 1996).

The majority of SNPs with L>>0 are unlikely to be recent mutations. At least 30% these of SNPs are present in both CEU and YRI populations with the same allele frequences. Only 7% of these SNPs are found in neither population, consistent with previous worldwide surveys of unselected variants (Int HapMap Consortium 2007) Further, a significant component (24%; L> 0.1) of these SNPs affect splicing through population-specific, frequency dependent selection. For these SNPs, the assumption of a common genetic landscape may not be appropriate.

## ACKNOWLEDGEMENTS

*Funding*: This research was undertaken thanks to support from the Canada Research Chairs program, the Canadian Foundation for Innovation, NSERC, and the University of Western Ontario.

## REFERENCES

Adami C. (1998) *Artificial Life*. p. 347. Springer.

Adami C, Ofria C, Collier TC. (2000) Evolution of biological complexity. *Proc Natl Acad Sci U S A*; **97**(9):4463-8.

Bershtein S, Segal M, Bekerman R, et al. (2006) Robustness-epistasis link shapes the fitness landscape of a randomly drifting protein *Nature* **444**:929-932,

Charlesworth J and Eyre-Walker A (2008) The McDonald-Kreitman test and slightly deleterious mutations. *Mol Biol Evol* **25**: 1007-15.

Crow JF. and Kimura M. (1970) *An introduction to population genetics theory*, Harper and Row.

Griffiths RC, Marjoram P. Ancestral inference from samples of DNA sequences with recombination. (1996) *J Comput Biol.* **3(4)**:479-502.

International HapMap Consortium. 2007 A second generation human haplotype map of over 3.1 million SNPs. *Nature*.**449**(7164):851-61.

Hirschorn JN. (2009) Genome-wide association studies--illuminating biologic pathways. *N Engl J Med.* **360**(17):1699-701.

Huang RS, Duan S, Bleibel WK, et al. (2007) A genome-wide approach to identify genetic variants that contribute to etoposide-induced cytotoxicity. *Proc Natl Acad Sci USA.*; **104**(23):9758-63.

Kimmel G, Karp RM, Jordan MI,and Halperin E. Association mapping and significance estimation via the coalescent. (2008) *Am. J. Hum. Genet.* **83**: 675-83.

Kodolitsch Yv. Berger J, Rogan PK. Predicting severity of haemophilia A and B splicing mutations by information analysis. (2006) *Haemophilia.***12**(3):258-62.

Kraft, P. and Hunter D.J. (2009) Genetic risk prediction: are we there yet? *N Engl J Med.* **360**(17):1701-1703.

Li CC. (1955) *Population genetics*. University of Chicago.

Maynard-Smith J. Natural Selection and the Concept of a Protein Space. *Nature* **225**, 563-564

Nalla VK and Rogan PK. (2005) Automated Splicing Mutation Analysis by Information Theory. *Hum Mut*; **25**:334-342.

Rogan PK and Schneider TD (1995) Using information content and base frequencies to distinguish mutations from genetic polymorphisms in splice junction recognition sites. *Hum Mutat.* **6**:74-76.

Rogan PK., Faux, B, and Schneider TD. (1998) Information analysis of human splice site mutations *Hum Mutat.* **12(3)**: 153-171.

Rogan PK. Svojanovsky S, and Leeder J. (2003) Information theory-based analysis of CYP219, CYP2D6 and CYP3A5 splicing mutations, *Pharmacogenetics*, **13(4)**: 207-218

Schneider TD. (1997) Information content of individual genetic sequences. *J Theor Biol.*;**189**(4):427-41